\documentclass[graybox]{svmult}

\usepackage{mathptmx}       
\usepackage{helvet}         
\usepackage{courier}        
\usepackage{type1cm}        
\usepackage{makeidx}         
\usepackage{graphicx}        
\usepackage{multicol}        
\usepackage[bottom]{footmisc}

\usepackage{dirtytalk}

\makeindex             

\begin{document}

\title*{Implementing the Cognition Level for Industry 4.0 by integrating Augmented Reality and Manufacturing Execution Systems}
\titlerunning{AR-based Cognition Level for Industry 4.0}

\author{Alfonso Di Pace and Giuseppe Fenza and Mariacristina Gallo and Vincenzo Loia and Aldo Meglio and Francesco Orciuoli}
\authorrunning{Di Pace A., Fenza G., Gallo M., Loia V., Meglio A., Orciuoli F.} 
\institute{Alfonso Di Pace, Giuseppe Fenza, Mariacristina Gallo, Vincenzo Loia, Aldo Meglio,\\ Francesco Orciuoli \at Dipartimento di Scienze Aziendali-Management and Innovation Systems,\\University of Salerno, Fisciano, Italy\\
\email{alfo.di.pace@gmail.com, gfenza@unisa.it, mgallo@unisa.it, loia@unisa.it,\\ aldo.meglio@gmail.com, forciuoli@unisa.it}
}
%
%
\maketitle

\abstract*{Each chapter should be preceded by an abstract (10--15 lines long) that summarizes the content. The abstract will appear \textit{online} at \url{www.SpringerLink.com} and be available with unrestricted access. This allows unregistered users to read the abstract as a teaser for the complete chapter. As a general rule the abstracts will not appear in the printed version of your book unless it is the style of your particular book or that of the series to which your book belongs.
Please use the 'starred' version of the new Springer \texttt{abstract} command for typesetting the text of the online abstracts (cf. source file of this chapter template \texttt{abstract}) and include them with the source files of your manuscript. Use the plain \texttt{abstract} command if the abstract is also to appear in the printed version of the book}

\abstract{
In the current industrial practices, the exponential growth in terms of availability and affordability of sensors, data acquisition systems, and computer networks forces factories to move toward implementing high integrating Cyber-Physical Systems (CPS) with production, logistics, and services. This transforms today’s factories into Industry 4.0 factories with significant economic potential. Industry 4.0, also known as the fourth Industrial Revolution, levers on the integration of cyber technologies, the Internet of Things, and Services.\\
This paper proposes an Augmented Reality (AR)-based system that creates a Cognition Level that integrates existent Manufacturing Execution Systems (MES) to CPS. The idea is to highlight the opportunities offered by AR technologies to CPS by describing an application scenario. The system, analyzed in a real factory, shows its capacity to integrate physical and digital worlds strongly. Furthermore, the conducted survey (based on the Situation Awareness Global Assessment Technique method) reveals significant advantages in terms of production monitoring, progress, and workers' Situation Awareness in general.\footnote{This is a post-peer-review, pre-copyedit version of an article published Barolli L., Amato F., Moscato F., Enokido T., Takizawa M. (eds) Advanced Information Networking and Applications. AINA 2020. Advances in Intelligent Systems and Computing, vol 1151. Springer, Cham. The final authenticated version is available online at: https://doi.org/10.1007/978-3-030-44041-1\_83}
}

\section{Introduction}
\label{sec:1}
The term \emph{Industry 4.0} has been used for the first time at the Hannover Fair in 2011. 
Industry 4.0 marks the dawn of the fourth Industrial Revolution through the use of cyber-physical systems (CPS), the Internet of Things, and Services.
The integration of cyber technologies that makes the products Internet-enabled, facilitates innovative services to achieve, among other things, Internet-based diagnostics, maintenance, operation, etc. in a cost-effective and efficient manner. Moreover, it helps the realization of new business models, operating concepts and smart controls, and focusing on the user and his/her individual needs~\cite{2}\cite{3}.

Cyber-Physical Systems (CPS) are defined as a transformative technology for managing interconnected systems between physical assets and computational capabilities~\cite{7}. Nowadays, higher availability and affordability of sensors, data acquisition systems, and computer networks forces factories to move toward implementing high-tech methodologies, to not suffer competition. 
In such an environment, CPS should be improved for managing Big Data and leveraging the interconnectivity of machines to make them more intelligent, resilient, and self-adaptable. Integrating CPS with production, logistics, and services can transform today’s factories into an Industry 4.0 factory with significant economic potential~\cite{7}.

\begin{figure}
\centering
\includegraphics[angle=0,width=0.9\linewidth]{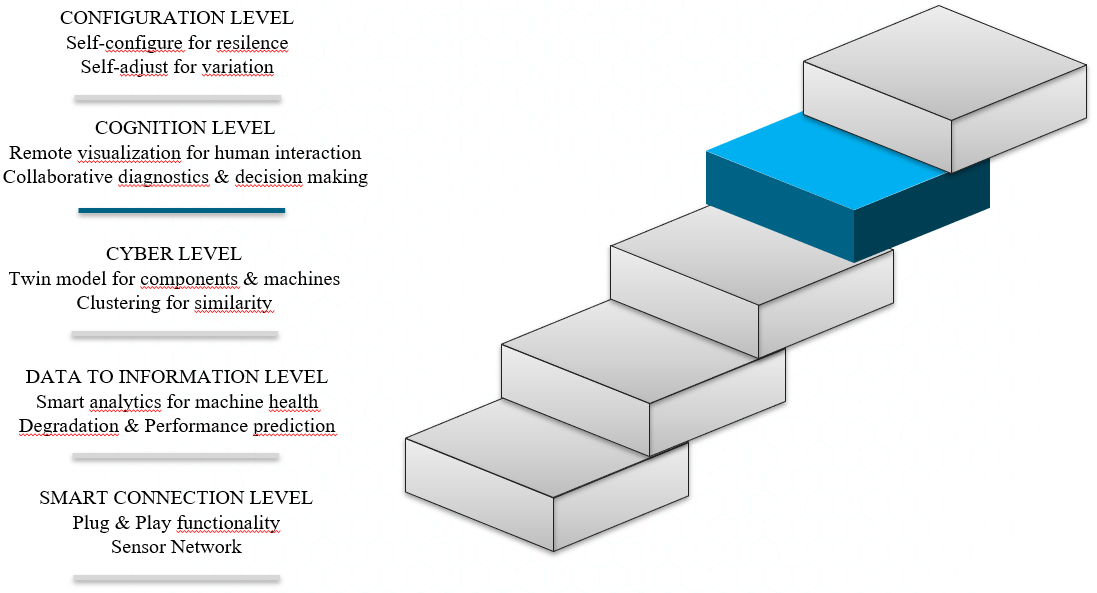}
\caption{Representation of CPS' levels.}\label{layers}
\end{figure}

The structure of CPS presents five levels (i.e., the 5C architecture): smart connection level, data-to-information
conversion level, cyber level, cognition level, and configuration
level (see Fig.~\ref{layers}). The 5C architecture provides a step-by-step guide for developing and deploying a CPS for manufacturing applications. In particular, the five layers define two main functional components: (1) the advanced connectivity that ensures real-time data acquisition from the physical world and information feedback from the cyberspace; and (2) intelligent data management, analytics and computational capability that constructs the cyberspace~\cite{17}.

Manufacturing Execution Systems (MES) are computerized systems used in manufacturing to track and document the transformation of raw materials to finished goods. MES operate in real-time across multiple function areas and jobs to enable the control of multiple elements of the production process (e.g., inputs, personnel, machines, and support services). So, they provide information helping decision-makers to understand how current conditions on the plant floor can be optimized to improve production output~\cite{1}.  By leveraging CPS and, in particular, by implementing a solution at the \emph{Cognition Level} of the 5C architecture, workers could have a better awareness of equipment status with less effort. This level will be responsible to collects both the individual machinery information and the integrated information from the machine network. The objective is having comprehensive knowledge about the whole monitored system~\cite{17}. 
The main focus is to define a model in which Augmented Reality (AR) acts as an intermediary between CPS and MES. 
AR refers to the integration of additional computer-generated content into a real-world environment~\cite{4,8}. Many current AR-based applications integrate computer graphics into the user's view of his current surroundings. AR applications have high potential to improve the user experience concerning the way users access and interact with information that has a direct spatial relation to the physical environment. Technological advances, especially concerning mobile devices, made AR-based apps plentiful and downloadable onto most mobile devices. 
However, their usage is limited mostly for entertainment and marketing applications. 

In the context of \emph{smart manufacturing}, this research work proposes a \emph{human-fabric interaction} by defining an AR-based model at the Cognition Level of CPS. In this model, the AR allows operators to exploit data, processed and provided by the MES, directly on the field during the task execution without hindering the operators'~actions. The advantage is to increase the operators' Situation Awareness (SA) and make the task execution more efficient and effective. \\

The remaining of this paper is structured as follows: Section~\ref{rw} illustrates the motivations driving our research and state of the art in this area. The proposed model is explained in Section~\ref{model}. Section~\ref{impl} details some implementation aspects, while Section~\ref{case_study} specifies a real case study. Finally, Section~\ref{conclusions} concludes the paper. 

\section{Motivation and Related Works}\label{rw}
In the area of Industry 4.0 and, in particular regarding the application of emerging technologies to \emph{smart manufacturing}, the main challenges are recognized:
\begin{itemize}
\item The increasing demand for new and ever more complicated products lead the manufacturers to address facilities with ever more sophisticated operations, resulting in more complex managing tasks also in terms of communication among equipment.
\item Even if a huge amount of data is collected, Big Data visualization and the interaction with the available knowledge is still complex to implement.
\item Even if CPS promise to smoothly include the human-in-the-loop for simplifying the manufacturing workflow, it is still necessary to ease human interaction and continuous monitoring.
\end{itemize}

One of the enabling technologies for Industry 4.0 is AR~\cite{15}. It could be useful, for instance, to reduce workers' travel within the factory and to leave their hands free from devices. There are different solutions for including AR in smart manufacturing. Usually, companies provide an AR-based software platform for industrial use in order to improve productivity, accuracy, quality, and safety of any skilled industrial workforce~\cite{16}. 

In the literature, Wang et al.~\cite{5} provide a survey about technical features, characteristics, and a broad range of applications of AR-based assembly systems. By analyzing papers from $1990$ to $2015$, authors conclude that the familiarity of developers and users with the scientific rationale and experience behind their applications strongly influences the effectiveness and usability of the new AR assembly systems. 

Fiorentino et al.~\cite{6} did an experiment where workers performed operations by using paper manuals with AR instructions. The results show that AR instructions the reduce error rate and improve performance in terms of the required time. 

In the context of \emph{smart manufacturing}, we are witnessing a clear evolution of the MES that enable coordination between the various functional areas (i.e., Production, Maintenance, Quality, and Logistics). However, the technological base needs a revision from agility and interoperability points of view. 
It should support the interaction between manufacturers and the monitoring platforms with low-latency~\cite{16}.


Concerning the AR solutions in Industry 4.0 recognized in literature, the idea underlying this paper is to build an augmented manufacturing environment employing data provided by the monitoring platform (e.g., MES solutions) to expand the physical world in which the manufacturers act. AR technologies represent the solution to achieve the proposed aim since they smoothly include human in the loop and provide \emph{human-fabric interaction}, giving to worker awareness about the state of the equipment.


\section{Model}\label{model}

Nowadays, the workers have to move inside the factory to monitor and control the facilities. There is not a direct interaction among equipment, and between equipment and MES system. To interact with the MES (i.e., read/insert data), the workers have to reach the facility physically, and 
to be aware about its state, 
they have to consult the MES panel (if present) or the documentation produced by the machines. Furthermore, in all these phases, the workers could have busy hands. 
Alternatively, the interaction between man and robot can happen virtually and remotely: the AR device transfers commands from the user to the robot as it was acting in the place. In this way, multiple systems can be controlled at the same time just staying everywhere in the factory (i.e., in a separate room, near another machine, and so on). 

Considering the huge amount of data that could cross the factory, together with AR technologies that can support the communication with MES, it is necessary to adopt Big Data technologies, in order to create a scalable system.  
While the AR can be used as a staff coordinator improving management and distribution of tasks, a real-time database enhances scalability and easily manages alerting scheme. These technologies realize a system that informs workers of their daily responsibilities and train them about the task previously completed by another worker.  
Through the use of the AR and smart devices, the workers have awareness about facilities wherever 
inside the factory. 
Efficiency is increased, and monitoring procedures are simplified, especially when some piece of equipment is difficult to reach physically. Furthermore, if the worker has problems with some processes, he can quickly consult manuals and receive help. With AR, workers will be able to use their devices to view a detailed floor plan with information on equipment location and tools.
\begin{figure}
\centering
\includegraphics[angle=0,width=0.8\linewidth]{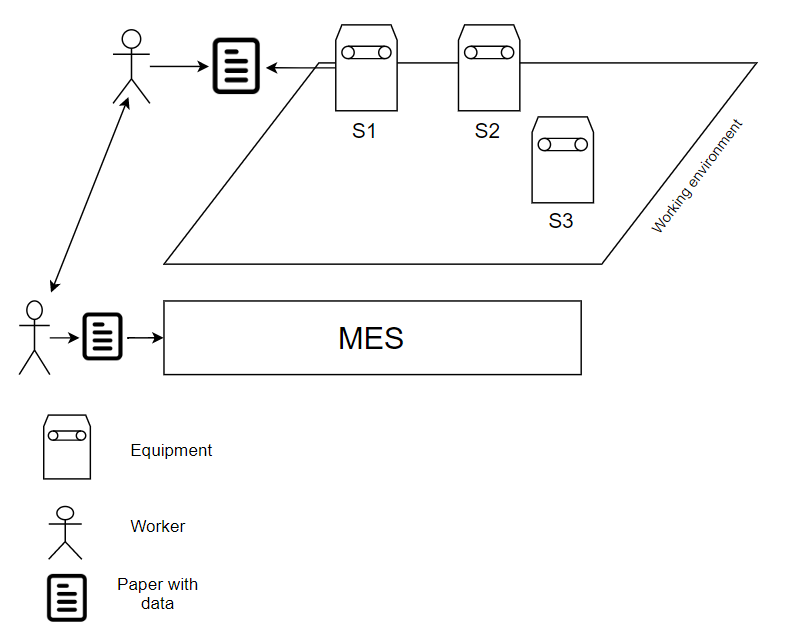}
\caption{Representation of the model without framework.}\label{without}
\end{figure}
\begin{figure}
\centering
\includegraphics[angle=0,width=0.8\linewidth]{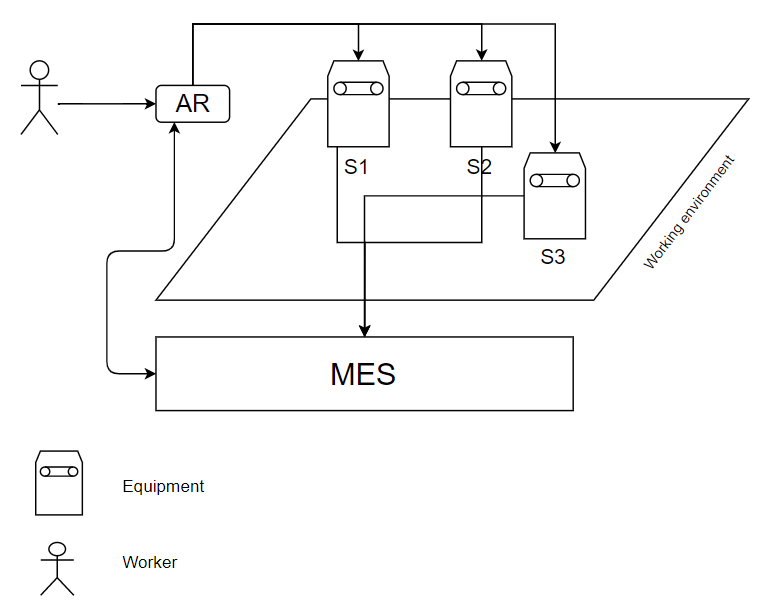}
\caption{Representation of the model described.}\label{with}
\end{figure}
The idea behind the proposed framework regards the necessity to build am AR-based system that can cooperate with the actual systems.
However, instead of only offering a User Interface (UI) that simplifies worker's tasks by monitoring in real-time the equipment state, we design a Cognition Layer that virtually connects CPS and MES (Fig.~\ref{without} and Fig.~\ref{with} show the differences between old systems and the proposed one). 
Building a framework that can operate at the Cognition level of the CPS, can offer a new way of managing equipment: the objective is to improve the decision-making capacity of the workers by giving them more situation awareness about the whole factory. 
Workers must be able to act as a supervisor and, if needed, react in real-time based on specific facility and specific process.

Definitely, our framework will offer the following functional features for the CPS:
\begin{itemize}
\item Decision making: designing a framework based on the human-in-the-loop model, make facilities' management faster and more flexible;
\item Human interaction: workers can visualize and interact with machines by a hologram through gestures that are transferred to the physical ones in real-time;
\item Visualization: integrating the information recorded by the MES and visualizing them supports previous features but, mostly, gives a clearer vision of the big picture (i.e., Situational Awareness for the workers).
\end{itemize}
Finally, the main challenges of this work can be summarized as follows: low latency in completing jobs, monitoring data in real-time, and integrate the Big Data information giving access to them from workers on the plant floor.

\section{Implementation Issues}\label{impl}
This section will expose technologies (hardware and software) involved during the implementation of the framework presented in the previous section.

Augmented Reality (AR) is a medium supported by a set of enabling technologies that are used to describe an environment where the physical world is enhanced by adding computer-generated objects. Using computer-vision methods makes them appear as if they coexist in the same dimension~\cite{13}. AR is provided to users through devices of different types known as AR devices. The main technologies supporting AR are related to:
\begin{itemize}
    \item Interaction: the user can interact with the augmented world by using gestures or physical input devices;
    \item Visualization: the user can use different devices classified as i) wearable technologies like head-mounted displays and smart glasses; ii) handheld displays like smartphones and tablets; iii) fixed projective displays like stationary screens and caves;
    \item Tracking: the task of detecting users’ position and orientation with respect to an environment to render digital objects by using the right perspective.
\end{itemize}
 The role of AR devices in the proposed methodology is twofold. Firstly, AR supports the workers involved in a maintenance task by delivering hints to them by considering different situations and different contexts. Secondly, AR devices can gather workers’ input and those features able to detect their situations and contexts.  Additionally, AR devices could also be empowered by additional types of sensors used to detect environmental features like humidity, temperature, etc. These further sensors could be both embedded into the AR device and deployed in the environment. All these inputs are useful for the cognitive system to incrementally learn and refine the model and also to generate contextualized hints.
In the AR field, tablets and smartphones were the first devices to offer this technology to everyone. Although their use results in a positive improvement, they could distract the workers, also busying their hands. Head-Mounted Display (HMD) devices overcome these issues by showing information in a hands-free manner and, simultaneously, allowing users' operations. In addition, the glasses can make use of gestures, touch, and voice recognition.

The physical devices used in this research are Microsoft Hololens\footnote{https://www.microsoft.com/it-it/hololens} v1. From the software point of view, we adopt Unity\footnote{https://unity.com/} and iMES\footnote{https://www.imes-solutions.com/en/root/industry-4-0-mes-solutions.html}. Data is stored in the cloud-hosted database Firebase\footnote{https://firebase.google.com/}.

Microsoft HoloLens is a pair of mixed reality smart glasses developed and manufactured by Microsoft. In this work, they are used to read data coming from the database and the MES and show information and helps.

Unity is a cross-platform game engine developed by Unity Technologies. Unity gives users the ability to create games and experiences in both 2D and 3D. In the presented work, Unity is the base of the framework, and it was used to implement the application.


IMES is an open-source manufacturing execution system application designed for Small-Midsize Job Shop Manufacturer. Employing an intelligent MES system with improved decision making can further enhance the performance of manufacturing and drive down production costs. In this work, all data pass through the iMES that communicate with the database and machines.


The Firebase Realtime Database is a cloud-hosted database. Data is stored as JSON and synchronized in real-time to every connected client. In this work, all data and MES' operations are stored inside Firebase.

\begin{figure}
\centering
\includegraphics[angle=0,width=0.8\linewidth]{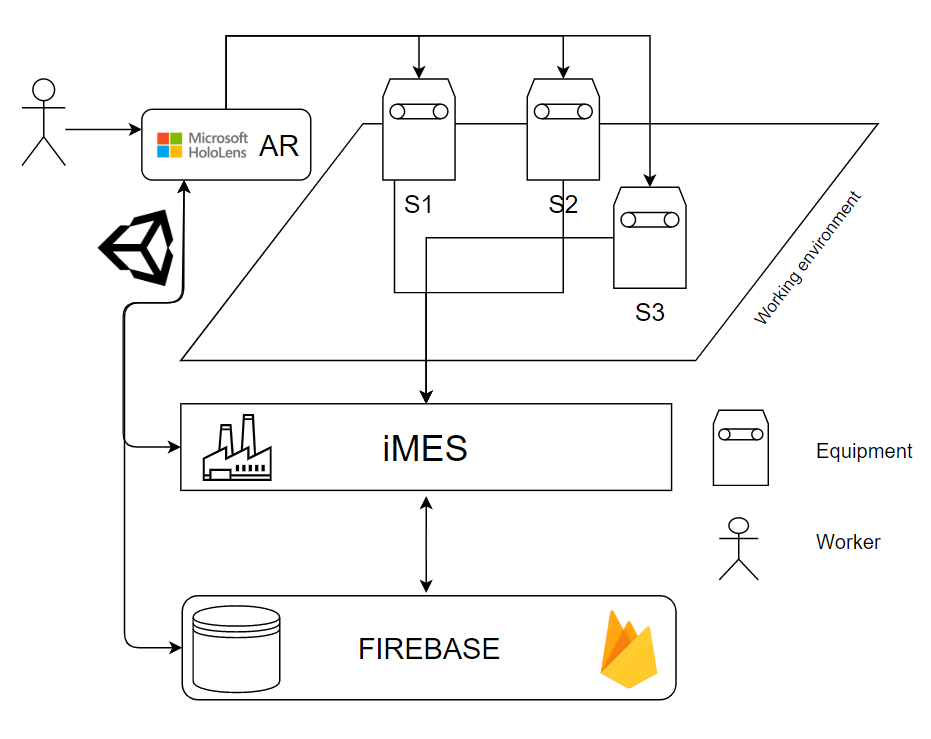}
\caption{Representation of the implementation architecture.}\label{visina8}
\end{figure}

\subparagraph{Implementation}

Implementation starts from iMes. The source code of the software was downloaded from the GitHub repository and built-in local, following documentation instructions. At this point, the built iMes  were pointing to a new Firebase database, that creates an account and set up a new factory. Firebase creates automatically all the entries associated with the account.

\begin{figure}
\centering
\includegraphics[width=12cm]{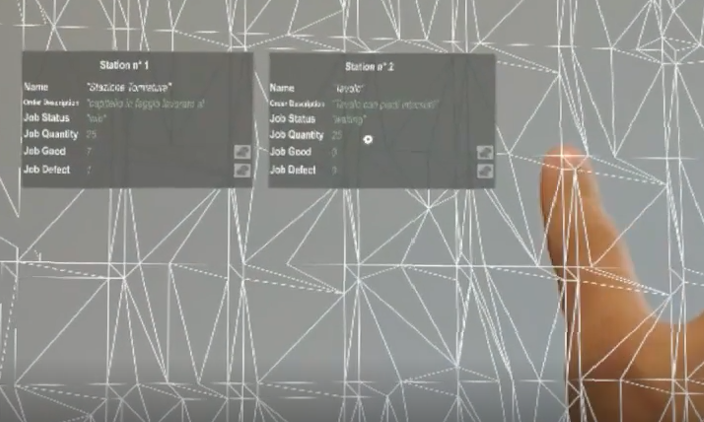}
\caption{Air-tap gesture to position a panel in the space.}\label{air-tap}
\end{figure}

Regarding Unity, in order to be compliant with Hololens, we adopt Holotoolkit\footnote{https://github.com/microsoft/MixedRealityToolkit-Unity}, the package distributed freely from Microsoft. 
The HoloToolkit package gives some immediate and basic function automatically, that can be added in the scene and in the project itself. Indeed, we were able to use the AR and IR camera to move an object into the scene using the pointer and the \emph{air-tap} gestures (an example is shown in Fig.~\ref{air-tap}. Furthermore, we use functionality that designates the object that has to follow our gaze. 
Finally, to make the menu always visible even rotating the head or to let the station panel always facing the user, the \emph{billboard} gives the chance to choose what axis is implied in the rotation.


\section{Case Study}\label{case_study}
In this section, we will expose an example of the real use of the realized system in order to demonstrate its efficiency and suitability for the application context. Furthermore, a formal evaluation of the system is discussed.



\subsection{Evaluation}
The scenario was based in a simulation on a typical workday in a real factory. Workers were invited to use the connected device 
during their work. They must use such a device to get information about facilities in the factory, reports about completed/rejected jobs, and turn on/off machinery. 
To test and evaluate the practical application of the system, we try it in a real factory by collaborating with two workers skilled in their tasks and in using facilities.

To have a clear spectrum of how the application could change the normal processes, the Situation Awareness Global Assessment Technique (SAGAT)~\cite{14} method has been adopted. It is usually used to measure individual SA in real-time simulations that support the human-in-the-loop model. Its output regards the capacity of the evaluated system to support operators' SA requirements. So, in this context, SAGAT helped to understand the usability of the UI, the level of completed tasks by the workers, and the level of general satisfaction and awareness of workers.

In particular, reached Situation Awareness is measured at three levels:
\begin{enumerate}
\item Level 1SA - Perception of the elements in the environment.\\
The first step in achieving SA is to perceive the status, attributes, and dynamics of relevant elements in the environment. The actual factory is an open and large space where facilities are distant from each other. Some of them require the use of a computer that is positioned near it; others require manual interaction from a user.
\item Level 2SA - Comprehension of the current situation.\\
It involves understanding the significance of objects and events in the environment and combining this data. Each worker has a workstation from which quickly reaching needed machinery. They will understand which machine is turned on, and is working.
\item Level 3SA - Projection of future status.\\
The highest level of situation awareness is to project the future actions of elements in the environment. The workers are able to understand where they have to go to execute their tasks. Also, they had a training course about their job and about safety.
\end{enumerate}

The SAGAT evaluation consists of simulating some real tasks and freezing the execution at various points along the time. In such pauses, workers ask questions designed to assess their level of SA. So, the answers are compared to the real situation, and then an interpretation will be done to provide an objective measure of SA.

In order to enable this type of evaluation, the proposed system was installed on a mobile device (e.g., a tablet), and integrate with the MES system. At this point, the application was able to show real-time information about some machines that were in the same production process. The test has been executed for an entire process, with two participants that have to do some tasks manually. Each question of the SAGAT method in the experiment was referred to each key step of the production process, about the safety and stress.

Regarding received answers, it emerges that workers, during the "Production process", through a tablet can have real-time information about other machinery physically far away, that can also coordinate remotely, having a more clear vision about all the tasks in the process. 

Since the application can alert when a machine is not working properly, we also ask questions about the safety aspects. We conducted a survey after no occurred event and after a simulated one. The answers showed that the workers can immediately recognize the alarm type and its source.

Finally, stress questions about workers' mood concerning the work environment, report that by working in a more supported way make workers feel at ease.\\

Definitely, the results of the simulation gave us good feedback about the usability of the application and the efficiency of the designed UI.  
Responses stated that the application makes the working environment comfortable in terms of requested skill level, reachability of facilities, and overall factory Situation Awareness.

\section{Conclusions}\label{conclusions}
This paper presents a system intending to create a Cognition Level between physical and virtual worlds in CPS. It leverages AR devices in order to assist and guide workers during their tasks. In particular, the framework implemented on Microsoft Hololens allows workers to be aware of facilities and their functioning/state. The application was experimented, with promising results, in a real factory by conducting a survey guided by the SAGAT model in order to test workers' awareness reached level. The right level of workers' SA results in low latency during the tasks, real-time monitoring, and workers' satisfaction.

In the future, some extensions could improve compatibility with other systems. For example, (1) managing vocal commands could give greater freedom of movement to the user; (2) adopting external library (e.g., Vuforia\footnote{https://library.vuforia.com/getting-started/overview.html}) could extend recognizable object in 2D or 3D, further improving the usability; (3) defining training function could help new worker to become familiar with the machinery and the tasks; (4) in-line help could guide users in application use.


\begin{acknowledgement}
This research was partially supported by:
\begin{itemize}
    \item the MIUR (Ministero dell'Istruzione dell'Università e della Ricerca) under the national program PON 2014-2020, Leonardo 4.0 (ID ARS01\_00945) research project in the area of  Industry 4.0
    \item the ECSEL-JU under the program ECSEL-Innovation Actions-2018 (ECSEL-IA) for research project CPS4EU (ID-826276) research project in the area Cyber-Physical Systems.
\end{itemize}

\end{acknowledgement}

 \bibliographystyle{plain}
 \bibliography{mybib}{}

\end{document}